\documentclass{rnaastex}

\begin{document}

\title{On the Pre-impact Orbital Evolution of 2018 LA, Parent Body of the Bright Fireball Observed over Botswana on 2018 June 2}

\correspondingauthor{Carlos~de~la~Fuente~Marcos}
\email{nbplanet@ucm.es}

\author[0000-0003-3894-8609]{Carlos~de~la~Fuente~Marcos}
\affiliation{Universidad Complutense de Madrid \\
             Ciudad Universitaria, E-28040 Madrid, Spain}

\author[0000-0002-5319-5716]{Ra\'ul~de~la~Fuente~Marcos}
\affiliation{AEGORA Research Group \\
             Facultad de Ciencias Matem\'aticas \\
             Universidad Complutense de Madrid \\
             Ciudad Universitaria, E-28040 Madrid, Spain}


\section{} 

On 2018 June 2, meteoroid 2018~LA became the third natural body ever to be observed before entering our 
atmosphere;\footnote{\href{https://minorplanetcenter.net/mpec/K18/K18L04.html}{MPEC 2018-L04: 2018 LA}} similarly small asteroids 2014~AA 
and 2008~TC$_{3}$ had stricken the Earth on 2014 January 2 \citep{2016Ap&SS.361..358D,2016Icar..274..327F} and 2008 October 7 
\citep{2009Natur.458..485J,2012P&SS...73...30O}, respectively. Asteroid 2008~TC$_{3}$ disintegrated as a superbolide over northern Sudan 
\citep{2009Natur.458..485J}, 19 hours after discovery, and 2018~LA completely broke up in a fireball over South Africa and Botswana, over 8 
hours after being first observed.\footnote{\href{https://cneos.jpl.nasa.gov/fireballs/}{Fireball and Bolide Data}} Meteoroid 2014~AA was not 
visually observed to disintegrate, but infrasound recordings showed that it exploded over the Atlantic Ocean less than a day after being 
first spotted \citep{2016Ap&SS.361..358D,2016Icar..274..327F}. Infrasound sensors also pinpointed the impact location of 2018~LA. The three 
asteroids had similar sizes of a few meters; such small bodies are probably fragments of larger objects. The study of the orbital evolution 
of such fragments prior to impact can help in understanding how asteroid disruptions take place \citep{2007MNRAS.382.1933T}.

The current orbit determination of 2018~LA (epoch JD~2458200.5, 23-March-2018, solution date 5-June-2018) is based on 14 observations for a 
data-arc span of 3.78~h and has semimajor axis, $a=1.374\pm0.002$~au, eccentricity, $e=0.4303\pm0.0009$, inclination, $i=4\fdg284\pm0\fdg004$, 
longitude of the ascending node, $\Omega=71\fdg8795\pm0\fdg0013$, and argument of perihelion, $\omega=256\fdg04\pm0\fdg03$; with an absolute 
magnitude of $30.6\pm0.3$, it may have been 2--5~m wide.\footnote{\href{http://ssd.jpl.nasa.gov/sbdb.cgi}{JPL's Small-Body Database}} This 
orbit determination is used to investigate the possible presence of known near-Earth objects (NEOs) moving in similar orbits, following the 
approach discussed by \citet{2015ApJ...812...26D,2016Ap&SS.361..358D} and applying the $D$-criteria. We found several objects with values of
the $D$-criteria with respect to 2018~LA under 0.05 and here we focus on three of them: (454100) 2013~BO$_{73}$ ($a=1.33192556\pm0.00000002$~au, 
$e=0.41841228\pm0.00000007$, $i=4\fdg543978\pm0\fdg000007$, $D_{\rm R}=0.024$), 2016~LR ($a=1.3813\pm0.0009$~au, $e=0.4351\pm0.0005$, 
$i=2\fdg542\pm0\fdg002$, $D_{\rm R}=0.0055$), and 2018~BA$_{5}$ ($a=1.3670\pm0.0009$~au, $e=0.4310\pm0.0005$, $i=4\fdg539\pm0\fdg007$, 
$D_{\rm R}=0.010$). The orbit of 2018~BA$_{5}$ is particularly close to that of 2018~LA in terms of $a$, $e$, and $i$. Our results are based 
on the latest orbit determinations available from JPL's Small-Body Database and our $N$-body simulations were performed as described by 
\citet{2015ApJ...812...26D,2016Ap&SS.361..358D}. These four objects exhibit rather chaotic orbital evolutions and experience flybys with 
Venus, the Earth--Moon system, and Mars.

\begin{figure}[!ht]
\begin{center}
\includegraphics[scale=0.4,angle=0]{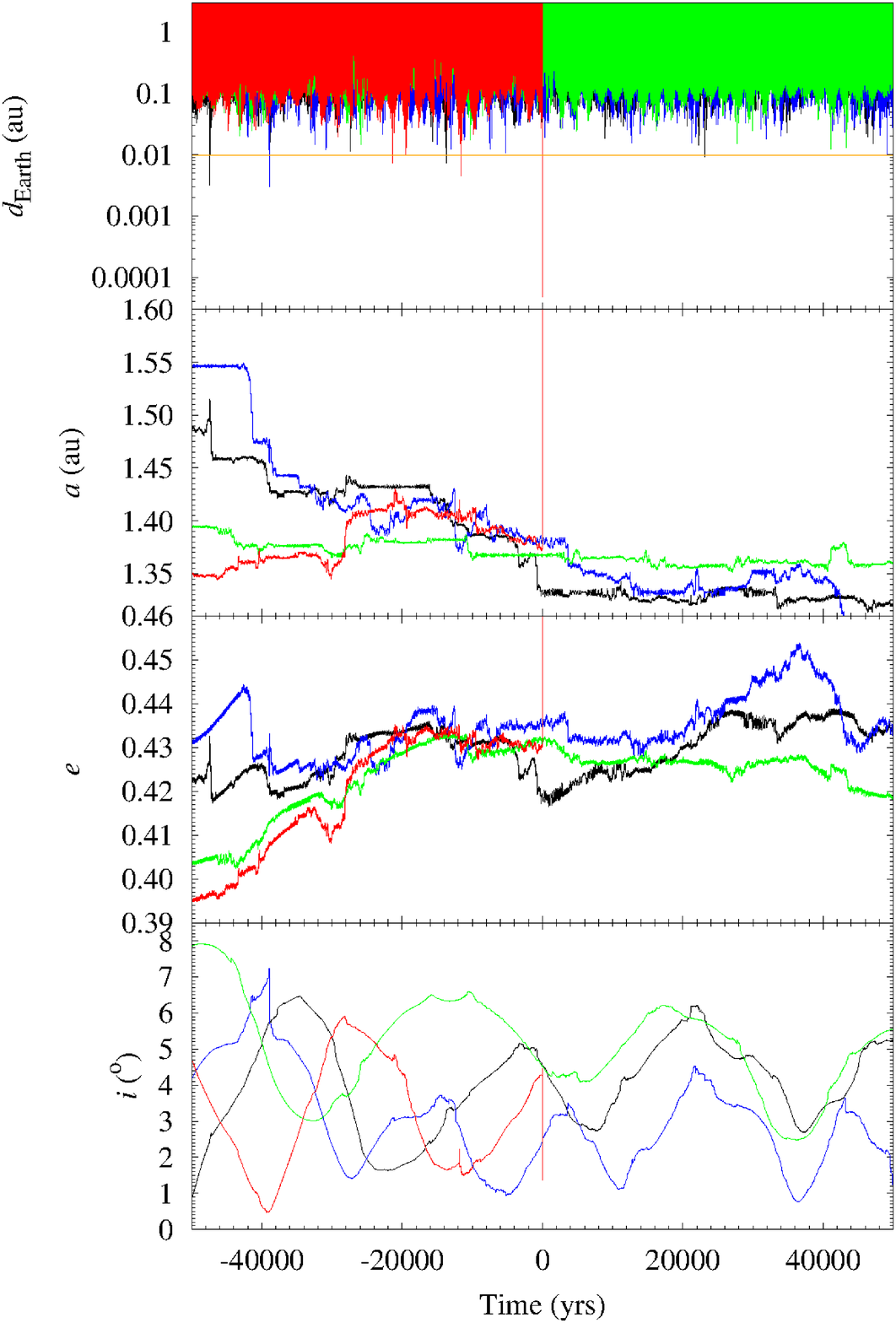}
\caption{Evolution of the values of the perigee distance (top panel) ---the Hill radius of the Earth is 0.0098~au (orange)--- semimajor axis 
         (second to top panel), eccentricity (second to bottom panel), and inclination (bottom panel) of the nominal orbits (zero instant of 
         time, epoch JD~2458200.5~TDB, 23-March-2018) of (454100) 2013~BO$_{73}$ (black), 2016~LR (blue), 2018~BA$_{5}$ (green), and 2018~LA 
         (red). Our calculations show that 2018~LA collided with our planet on JD~2458272.20259~TDB, JPL's Small-Body Database shows as impact 
         time JD~2458272.19814~TDB.  
\label{fig:1}}
\end{center}
\end{figure}

Figure~\ref{fig:1} shows the pre-impact orbital evolution of 2018~LA and the three NEOs moving along relatively similar orbits. All of them 
encounter the Earth at distances close to or below the Hill radius (top panel). The evolution in $a$ (second to top panel) and $e$ (second 
to bottom panel) is somewhat alike (at least during part of the simulated time), but that of $i$ (bottom panel) is less correlated. 
These objects cannot be the result of a recent fragmentation event, but they could be part of a dynamical grouping \citep{2016MNRAS.456.2946D}.
 
In this Note, we have explored the pre-impact orbital evolution of 2018~LA, the parent body of the fireball observed over South Africa and
Botswana on 2018 June 2. We have identified a few known NEOs with orbits similar to that of the impactor. Although the overall dynamical
evolution of these NEOs bears some resemblance, it cannot be concluded that any of them might have been physically related to each other in 
the relatively recent past (see Figure~\ref{fig:1}). On the other hand, three other bolides were observed early in June: Crete on 2002 June 
6, Washington State on 2004 June 3, and Reisadalen on 2007 June 7 \citep{2015MNRAS.446L..31D}. It is unclear whether these events could be 
related to each other. An even brighter superbolide was observed over Botswana on 2009 November 21 \citep{2015MNRAS.446L..31D}.

\acknowledgments

We thank S.~J. Aarseth for providing the code used in this research and A.~I. G\'omez de Castro for providing access to computing facilities. 
This work was partially supported by the Spanish MINECO under grant ESP2015-68908-R. In preparation of this Note, we made use of the NASA 
Astrophysics Data System and the MPC data server.

\end{document}